\documentstyle[epsf,rotate,12pt]{article}

\newcommand{\fsl}{\hskip -2mm /}
\newcommand{\fslsmall}{\hskip -1.5mm /}
\newcommand{\Nfat}{ {\rm I\hspace{-0.23em}N} }
\newcommand{\rar}{\rightarrow}
\newcommand{\beq}{\begin{equation}}
\newcommand{\eeq}{\end{equation}}

\begin{document}

\begin{flushright}
TUM-T31-93/95 \\
hep-ph/9512252 \\
March 1996 \\
(revised version) \\
\end{flushright}
\begin{center}
\vskip 2cm
\Large
{\bf Bremsstrahlung corrections to the decay $b \rightarrow s \gamma$} 
\vskip 1.5cm
\large
Nicolas Pott
\vskip 1cm
\small
{\em Physik-Department, Technische Universit\"at M\"unchen \\
 D-85747 Garching, Germany}

\end{center}
\vskip 2cm
\begin{abstract}
We calculate the O($\alpha_s$) gluon Bremsstrahlung corrections to the
inclusive decay $b \rightarrow s \gamma$, involving the full operator
basis $\hat O_1$ -- $\hat O_8$. Confirming and extending earlier
calculations of Ali and Greub, we give formulas for the total decay
width as well as the perturbative photon spectrum, regarding the
former as a necessary part of the forthcoming complete NLO
analysis. We explore in detail the renormalization scale dependence of
our results and find it considerably increased.
\end{abstract}
\thispagestyle{empty}

\newpage
\pagenumbering{arabic}

\section{Introduction}

If we define rare weak decays of hadrons as weak decays which are rare
because they are loop-induced (as opposed to CKM-suppressed), the
analysis of these decays may shed light on at least three important
topics: (i) Since they can proceed even at leading order only through
diagrams with a loop of virtual particles, these decays do test the
standard model of weak interactions as {\em quantum} field theory.
(ii) Since they are decays of hadrons, they are of course strongly
affected by QCD effects. If, however, one naively calculated these
corrections, due to the massiveness of the vector bosons in the loop
one would have to cope with large logarithms $\ln {M_W \over m_q}$,
where $M_W$ is the electroweak scale and $m_q$ the usual hadronic
scale. These large logarithms spoil perturbation theory: one has to
resum them invoking the powerful techniques of the renormalization
group. Especially in the case $ b \rightarrow s \gamma$, where the
strong interaction is known to enhance the decay rate by a factor of
2-3, we have hence an ideal testing ground for those frequently
employed concepts of partial resummation of the perturbative
series. (iii) Since they are rare even in the standard model, they are
sensitive to the effects of new physics: plainly spoken, new heavy
particles running in the loop will contribute to the decay if their
masses are not much larger than the electroweak scale, as expected for
supersymmetry, left-right-symmetric models, technicolor and so on. Of
course also the couplings of these particles must not be too small: so
if we can find no positive sign of new physics, i.e.\ a deviation from
the Standard Model (SM) prediction, some more or less sharp
restrictions on the parameter space are all we can hope for.

Among all rare weak decays $b \rar s \gamma$ takes a special role: it
is of order $G_F^2 \alpha$ (not $G_F^4$ or $G_F^2 \alpha^2$), and
accordingly the corresponding branching ratio is much larger than that
of most other rare decays. In fact, within the field of $B$ physics,
$b \rar s \gamma$ is the only one which has been measured
experimentally, and this only recently. Following the first
observation of the exclusive mode $B \rightarrow K^{\ast} \gamma$ in
1993 \cite{cleo1}, the CLEO collaboration reported by now measurements
\cite{cleo2} of the inclusive branching ratio, ${\rm BR}[B \rar X_s
\gamma] = (2.32 \pm 0.67) \times 10^{-4}$ as
well as of the photon spectrum. Here $X_s$ denotes an arbitrary state
of total strangeness $-1$, and experimentally some lower cutoff in the
photon energy has to be imposed in order to exclude the tree-level
channel $b \rar c \bar c s \gamma$.  As it is well known, the
inclusive rate is of much more theoretical interest than the exclusive
modes, because within the framework of Heavy Quark Effective Theory
(HQET) one can show that it is given by the free quark decay model
plus perturbative QCD, up to calculable corrections of order $1/
m_b^2$ \cite{HQET}. The same is surprisingly true for suitable defined
moments of the photon spectrum \cite{neubert}. Recall that a
continuous photon spectrum in $b \rar s \gamma$ arises by reason of
two effects: perturbatively, by emission of gluon bremsstrahlung (the
topic of our paper), which results in a long tail of the spectrum
below the kinematical endpoint, and non-perturbatively by the Fermi
motion of the $b$ quark inside the $B$ meson, which leads to a
symmetric smearing around the endpoint. Combining these two effects is
a quite non-trivial task, primarily because one has to make a
consistent separation between the perturbative and the
non-perturbative region. The analysis of the spectrum has been the
subject of some recent papers \cite{shifman, ali1, kapustin,
kapustin2}, and it will thus not be the central point of our work;
instead we will give {\em complete} results for the bremsstrahlung
contribution to the total decay width. To our knowledge, there exists
only one calculation of Ali and Greub \cite{ali2} on this subject,
which for a long time was in need of an independent check. In view of
the increasing desire for precise predictions of ${\rm BR}[B \rar X_s
\gamma]$ we have performed this check, confirming and extending
Ref.\ \cite{ali2}. 

Our paper is organized as follows: In section 2 we give the effective
Hamiltonian which we used in our calculations, and discuss very
briefly the structure of a complete Next-to-Leading-Order (NLO)
analysis and the place of this paper within that greater task. In
section 3 we give the amplitude $b \rar s \gamma g$ using this
Hamiltonian, and we also calculate those O($\alpha_s$) corrections to
the amplitude $b \rar s \gamma$ which are needed for cancellation of
infrared divergences. In section 4 we compute photon spectrum and
total decay width from this amplitude, giving some details about the
phase space integration in $D=4-2\epsilon$ dimensions. Subsequently,
in section 5 we analyze these results numerically, exploring their
range of applicability and discussing their dependence on the
renormalization scale as well as on the other input parameters. The
paper closes with a short summary and outlook in section 6.

\section{The effective Hamiltonian}

In order to make use of the renormalization group techniques for
calculation of short-distance QCD effects, we work within the
framework of an effective five quark theory where the W boson and the
top quark have been removed as explicit dynamical degrees of
freedom. Neglecting contributions with smaller CKM parameters ($ \vert
V_{ub} V_{us}^{\ast} \vert / \vert V_{tb} V_{ts}^{\ast} \vert < 0.02
$), the relevant Hamiltonian for the process $b \rar s \gamma, s
\gamma g$ is in leading order of the operator product expansion given
by \cite{grinstein}
\beq \label{hamiltonian}
\hat H_{eff}=-{4 G_F \over \sqrt{2}}V_{tb}V_{ts}^{\ast}
\sum\limits_{i=1}^{8}
C_i(\mu) \hat O_i.
\eeq
Here, $G_F$ is the Fermi coupling constant, $V_ {ij}$ are the CKM
matrix elements, $C_i(\mu)$ denote the Wilson coefficients evaluated
at the scale $\mu$, and the operator basis reads
\begin{eqnarray}
\nonumber
& &\hat O_1 = (\bar c_{L \beta} \gamma^{\mu} b_{L \alpha}) (\bar s_{L
\alpha}
\gamma_{\mu}c_{L \beta}), \\
\nonumber
& &\hat O_2 = (\bar c_{L \alpha} \gamma^{\mu} b_{L \alpha}) (\bar s_{L
\beta}
\gamma_{\mu} c_{L \beta}), \\
\nonumber
& &\hat O_3 = (\bar s_{L \alpha} \gamma^{\mu} b_{L \alpha})
\sum\limits_{q=u,d,s,c,b}
(\bar q_{L \beta} \gamma_{\mu} q_{L \beta}), \\
\nonumber
& &\hat O_4 = (\bar s_{L \alpha} \gamma^{\mu} b_{L \beta})
\sum\limits_{q=u,d,s,c,b} ( \bar q_{L \beta} \gamma_{\mu} q_{L
\alpha}), \\
\nonumber
& &\hat O_5 = (\bar s_{L \alpha} \gamma^{\mu} b_{L \alpha})
\sum\limits_{q=u,d,s,c,b} ( \bar q_{R \beta} \gamma_{\mu} q_{R \beta}),
\\
\nonumber
& &\hat O_6 = (\bar s_{L \alpha} \gamma^{\mu} b_{L \beta})
\sum\limits_{q=u,d,s,c,b} ( \bar q_{R \beta} \gamma_{\mu} q_{R
\alpha}), \\
\nonumber
& &\hat O_7 = {e \over 16 \pi^2} \bar s_{\alpha} \sigma^{\mu \nu} (m_b
R + m_s L) b_{\alpha} F_{\mu \nu}, \\ & &\hat O_8 = {g_s \over 16
\pi^2} \bar s_{\beta} \sigma^{\mu \nu} (m_b R + m_s L) T_{\beta
\alpha}^a b_{\alpha} G_{\mu \nu}^{a},
\end{eqnarray}
 
with $e$ and $g_s$ denoting the electromagnetic and strong coupling,
respectively, and the projection operators $R,L = {1 \over 2} (1 \pm
\gamma_5)$.
Some caution is necessary if we apply this result, which was
originally obtained only for the process $b \rar s \gamma$, to the
process $b \rar s \gamma g$. This is, because potentially new diagrams
like the one shown in Fig.\ 1
\begin{figure}[hbt]
\centerline{
\epsffile{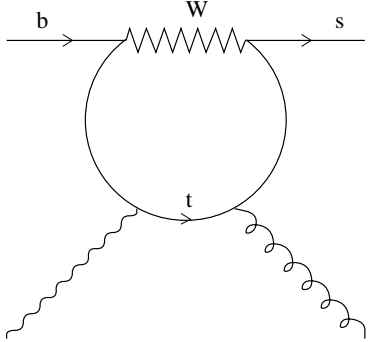}
}
\vspace{0cm}
\caption{\small A standard model Feynman diagram contributing to the
decay $b \rar s \gamma g$.  Diagrams of this type lead to low energy
effective operators which are not contained in the Hamiltonian (1),
which was originally derived for the process $b \rar s \gamma$
only. However, these operators are of dimension 8 or higher and hence
are suppressed by powers of $m_b^2/m_W^2$.}
\label{smdiagram}
\end{figure}
could lead to new operators containing two quark fields, the photon
and the gluon field simultaneously. However, a detailed analysis
following the lines of \cite{grinstein}, i.e.\ using gauge invariance
and the equations of motion, shows that all these operators must
necessarily be of dimension 8 and hence are suppressed by powers of
$m_b^2/m_W^2$.

The inclusive decay rate was calculated in the approximation of
leading logarithms, i.e.\ holding terms of order $(\alpha_s \ln M_W /
m_b)^n, n=0,...,\infty$. An exhaustive discussion of these results can
be found in Ref.\ \cite{LO}. In this reference, there is also analyzed
the general structure of next-to-leading calculations, which hold
terms of order $\alpha_s (\alpha_s \ln M_W / m_b)^n,
n=0,...,\infty$. The main points we would like to stress here are: (i)
The dominant theoretical uncertainty in the leading order result
arises from the renormalization scale dependence of the Wilson
coefficient $C_7(\mu)$. It amounts to about $ \pm 25\%$ \cite{LOali},
and it can hopefully be reduced by the full NLO calculation to about $
\pm 3\%$.  (ii) The full NLO calculation requires three steps:
O($\alpha_s$) matching of the effective theory to the SM matrix
elements at $\mu=M_W$ which gives $C_i(\mu=M_W)$, evolution of these
Wilson coefficients down to $\mu \simeq m_b$ with the help of the NLO
anomalous dimension matrix, calculation of O($\alpha_s$) corrections
to the matrix elements of the effective theory at the low scale.

From these three steps the first has been tackled in Ref.\ \cite{adel}.
Certain parts of the second have also been calculated, see Ref.\
\cite{ciuchini} for a summary of the present state of the
art. Unfortunately, the most difficult (and presumably fairly
significant) ones, involving finite parts of two-loop and divergent
parts of three-loop diagrams, are not yet known. The last step splits
into two pieces: O($\alpha_s$) corrections to the $b \rar s \gamma$
amplitude, and the bremsstrahlung amplitude $b \rar s \gamma g$ to
O($\alpha_s$). The latter has to be considered, because the final
state $s$+gluon contributes to the measured state $X_s$ as well as a
single strange quark. In the following sections, we calculate these
bremsstrahlung corrections using the full operator basis (2). We would
like to emphasize that the contributions of the penguin operators
$\hat O_3$ -- $\hat O_6$, which were neglected in \cite{ali2}, should
not be considered numerically negligible to begin with although their
Wilson coefficients are small: in leading order, their effect on the
decay width amounts to about 15\% (in the NDR scheme\footnote{Unless
noted otherwise, by {\em scheme} we mean throughout this paper the
regularization scheme used for the treatment of $\gamma_5$ in $D$
dimensions. NDR stands for Naive Dimensional Regularization, and HV
for the 't~Hooft-Veltman scheme.}), hence a consistent calculation
certainly should include all their contributions beyond leading order,
too.

We now proceed to give a somewhat detailed calculation of the above
mentioned brems-strahlung contributions, closing with some numerical
results at the end of section 5.

\section{The amplitude $b \rightarrow s \gamma g$}

Using the effective Hamiltonian (1) we will calculate in this section
the complete transition amplitude for the decay
\begin{eqnarray*}
& & b_{\alpha}(p) \rightarrow s_{\beta}(p') + \gamma(q, \epsilon) + g_a(r, \eta),
\end{eqnarray*}
where $\alpha$, $\beta$ and $a$ denote colour indices and $\epsilon$
and $\eta$ are the four dimensional polarization vectors of photon and
gluon, respectively.  The amplitude will have the form
\begin{equation}
M^{brems}= \sum\limits_{i=1}^8 M_i = T^a_{\beta \alpha} V
\sum\limits_{i=1}^8 \tilde 
C_i(\mu) 
 \bar u(p') T_i u(p), \label{brems}
\end{equation}
where we have extracted the couplings in a common factor
\begin{equation}
V={- i G_F \over \sqrt{2} \pi^2} V_{tb} V_{ts}^{\ast} g_s e, \label{V}
\end{equation}
$T^a$ denotes as usually the Gell-Mann matrices, $T_i$ are some Dirac
structures depending on $p$,$q$,$r$, $\epsilon$ and $\eta$, and
$\tilde C_i$ are the so-called effective Wilson coefficients
\cite{LO} which are linear combinations of the usual Wilson
coefficients (\ref{hamiltonian}), defined (in the NDR scheme) by
\beq \label{effwilson}
\tilde C_{1 \ldots 6} = C_{1 \ldots 6}, \; \; \tilde C_7 = C_7 - {1 \over 3}
C_5 - C_6, \; \;
\tilde C_8 = C_8 + C_5.
\eeq
In the HV scheme, $\tilde C_i = C_i$. The reason for introducing these
quantities will be explained after calculating the matrix elements of
the penguin operators $\hat O_3 -
\hat O_6$ in Sec.\ 3.3.

Throughout this section, we work (where necessary) in $D=4 - 2
\epsilon$ dimensions in order to regularize UV divergences, and we
treat $\gamma_5$ in $D$ dimensions according to the NDR scheme.

\subsection{Magnetic penguins: $\hat O_7$ and $\hat O_8$}
The magnetic penguin operators $\hat O_7$ and $\hat O_8$ contribute
only through the tree-level diagrams of Fig.\ 2
\begin{figure}[t]
\centerline{
\epsffile{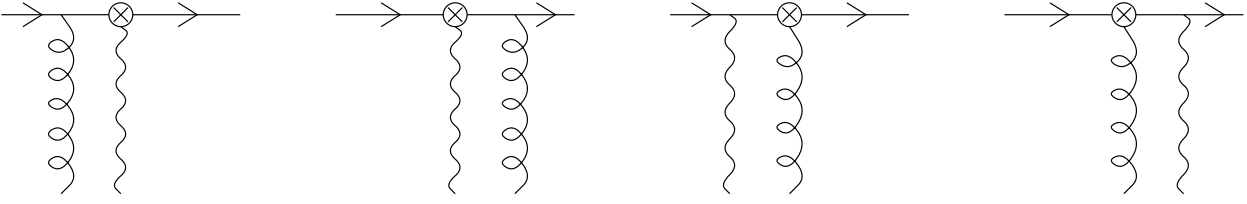}
}
\vspace{0cm}
\caption{\small The tree-level Feynman diagrams which contribute within the
effective theory (1) to the decay $b \rar s \gamma g$.}
\label{treelevel}
\end{figure}
which yield trivially
\begin{eqnarray}
T_7 &=& {1 \over 2}  \left [ {1 \over -2 p r }(m_b R+m_s L)q \fsl 
\epsilon \fsl (p \fsl - r \fsl + m_b)
 \eta \fsl + {1 \over 2 p'r} \eta \fsl ( p \fsl-q \fsl + m_s) q \fsl
  \epsilon \fsl (m_b R+m_s L)
\right ], \label{t7} \\
T_8 &=& {Q_d \over 2} \left [ {1 \over -2 p q } (m_b R + m_s L) r \fsl
\eta \fsl (p \fsl - q\fsl + m_b) \epsilon \fsl + {1 \over 2 p'q}
\epsilon \fsl ( p
\fsl - r \fsl + m_s) r \fsl \eta \fsl (m_b R + m_s L) \right ],
\label{t8}
\end{eqnarray}
with $Q_d = -1/3$.

Now these parts of the bremsstrahlung amplitude stemming from the
magnetic penguins are affected by infrared divergences. That is, after
phase space integration $\vert M_7 \vert ^2$ and $\vert M_8 \vert ^2$
yield no finite decay rate. Therefore the first step is to regularize
these divergences: for the sake of simplicity, we choose dimensional
regularization \cite{sirlin} so that the phase space integration is to
be done in $D = 4 - 2\epsilon$ dimensions and therefore becomes a
little bit involved. The second step is to eliminate the
divergences. According to the well-known Bloch-Nordsieck mechanism
\cite{blochnordsieck}, this can be achieved by considering the
appropriate diagrams with virtual instead of real gluons. Since the
infrared divergences of $\hat O_8$ occur only at the low energy
endpoint of the photon spectrum which is experimentally not
accessible, they are only of minor interest and need not to be
considered here. We are left with the one diagram of Fig.\ 3f which
yields (together with the $O(\alpha_s^0)$ contribution) in the Feynman
gauge and before renormalization:
\beq
M_7^{virt} =  
{i e G_F \over 2 \sqrt{2} \pi^2} V_{tb} V_{ts}^{\ast} C_7(\mu)
\delta_{\alpha \beta} \bar
u(p') T_{7,virt}^{(0)} u(p), \label{virt}
\eeq
with
\begin{eqnarray}
\nonumber
& &T_{7,virt}^{(0)} = (m_b R + m_s L) \epsilon \fsl q
\fsl (1 + K_g^{(0)}),
\\
\nonumber
& &K_g^{(0)} = {\alpha_s \over 3 \pi} (4 \pi)^{\epsilon} \Gamma
(1+\epsilon) {1+\rho \over 1-\rho} \left [ - {1 \over \epsilon} \ln \rho + {1
\over 2} \ln ^2 {m_s^2 \over \mu^2} - {1 \over 2} \ln ^2 {m_b^2 \over
\mu^2} - 2 \ln \rho \right ], \\ & & \rho = {m_s^2 \over
m_b^2}. \label{virt2}
\end{eqnarray}
Note that this amplitude is UV finite: the $1/\epsilon$ poles
correspond to infrared singularities. If we wish to calculate the
transition amplitude, we also must make allowance for diagrams with
quark self energy parts. To this end, remember that according to the
LSZ reduction formula the transition amplitude is given by the
corresponding truncated Green function times the residua of the
propagators of the external particles to the same order in
$\alpha_s$. Now let us employ the on-shell renormalization condition
\beq
{\partial \Sigma(p) \over \partial p \fsl} \vert_{p \fslsmall = m} = 0
\eeq
for determining the finite parts of the wave function renormalization
constant $Z_\psi$, which yields
\beq
Z_{\psi}(m) = 1 - {\alpha_s \over 3 \pi} \left ( {3 \over \epsilon} -
3
\gamma + 3 \ln {4 \pi \mu^2 \over m^2} + 4 \right )
\eeq
in Feynman gauge (here the singularity contains an $1/\epsilon$ UV and
an $2/\epsilon$ IR contribution). With this choice, the residuum of
the renormalized quark propagator equals 1 (to one loop), and the
transition amplitude is simply given by the renormalized truncated
Green function, i.e.\ by the amplitude (\ref{virt}) given above after
including renormalization. To achieve the latter, we have to add the
counterterm amplitude $(Z_{77} Z_m Z_{\psi}^{1/2}(m_b)
Z_{\psi}^{1/2} (m_s) - 1) M_7^{virt}$, where we choose the operator
and mass renormalization constants $Z_{77}$ and $Z_m$ according to
the $\overline {MS}$ scheme \cite{msbar}.  At this place, employing
the $\overline {MS}$ scheme is of course necessary, because the whole
calculation of the Wilson coefficients has been done in this scheme.
The result of this renormalization procedure amounts to replacing
$T_{7,virt}^{(0)}$ in (\ref{virt}) by
\beq \label{t7virtR}
T_{7,virt}^{(R)} = (\overline m_b(\mu)R +\overline m_s(\mu)L) \epsilon
\fsl q \fsl (1+ K_g^{(R)}),
\eeq
with 
\begin{eqnarray}
\nonumber
K_g^{(R)} \; = & & {\alpha_s \over 3 \pi} (4 \pi)^{\epsilon}
\Gamma(1+\epsilon) \left \{ - { 1 \over \epsilon}
\left [ {1+\rho \over 1-\rho} \ln \rho + 2 \right ] \right. \\ 
& & \left.  + {1+\rho \over 1-\rho } \left
[ {1 \over 2} \ln ^2 {m_s^2 \over
\mu^2} -
{1 \over 2} \ln ^2 {m_b^2 \over \mu^2}  - 2   \ln \rho
\right ] + {3 \over 2} \ln {m_s^2 \over \mu^2}
+ {3 \over 2} \ln {m_b^2 \over 
\mu^2} - 4 \right \} \label{kgR} 
\end{eqnarray}
and $\overline m_q(\mu)$ the running quark masses in the
$\overline{MS}$ scheme.  Again, this is UV finite.

\subsection{Current-current operators: $\hat O_1$ and $\hat O_2$}

The current-current operators contribute through diagrams of the type
shown in Fig.\ 3a.
\begin{figure}[t]
\centerline{
\epsfysize=11cm
\epsffile{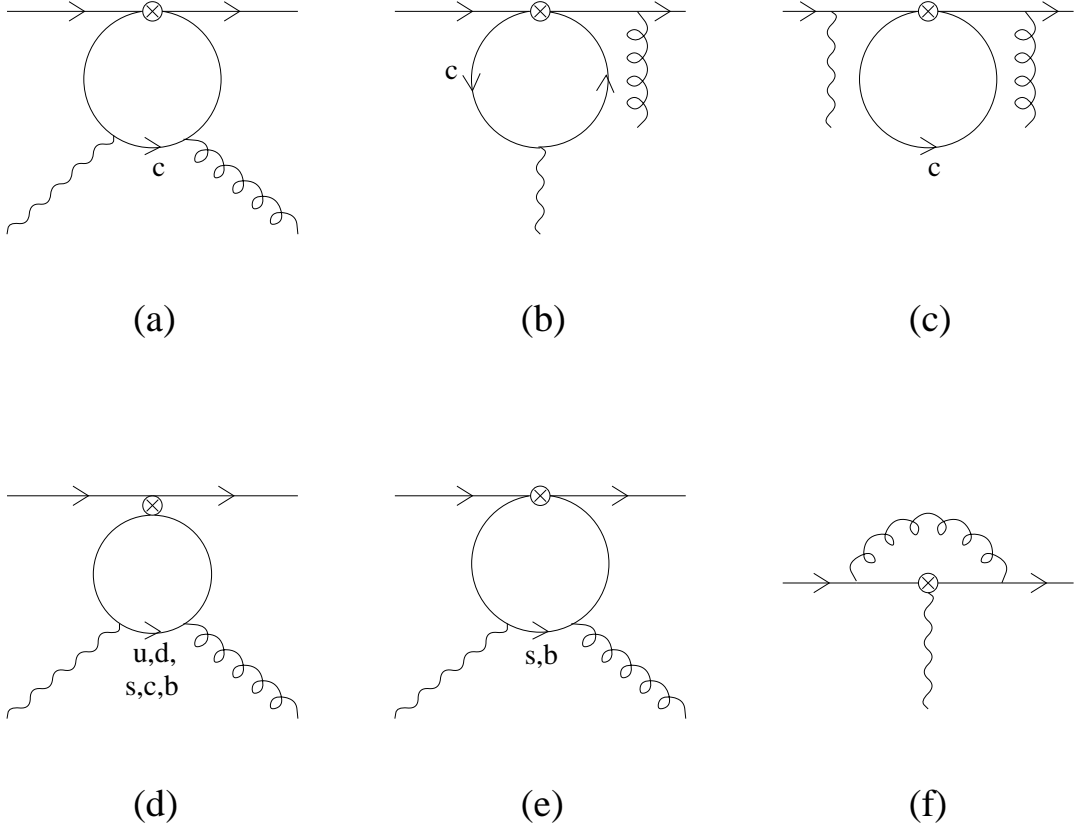}
}
\vspace{0cm}
\caption{\small Examples of one-loop Feynman diagrams which contribute
within the effective theory (1) to the decay $b \rar s \gamma g$. (a),
(b) and (c) depict insertions of the current-current operators,
whereas (d) and (e) illustrate the two distinct types of insertions of
penguin operators. (f) displays the virtual gluon correction to the $b
\rar s \gamma$ matrix element of the magnetic penguin operator $\hat
O_7$. This graph is needed for cancellation of infrared divergences at
the high energy endpoint of the photon spectrum in $b \rar s \gamma g$.}
\label{feyndiagram}
\end{figure}
Note that diagrams as Fig.\ 3b in which only one particle is radiated
from the loop vanish on-shell, as one may easily verify through
explicit calculation. From chirality, the same is true for diagrams as
Fig.\ 3c in which there is no particle at all radiated from the loop.

Since their colour structure entails a factor ${\rm tr}\,T^a$, all
diagrams with an insertion of $\hat O_1$ vanish:
\beq
T_1 = 0. \label{t1}
\eeq
Using some easily calculable one-loop three-point functions, one
obtains for the diagrams involving $\hat O_2$ 
\beq
T_2 = Q_u \kappa \left ( { m_b^2 \over m_c^2} s \right ) W_2, \label{t2}
\eeq
with
\begin{eqnarray}
& &\hspace*{-1cm}W_2 =  \left \{ {1 \over q r} \left [
(\eta q) q \fsl \epsilon \fsl r \fsl - (\epsilon r) q \fsl \eta \fsl r
\fsl
 - (\epsilon r) (\eta q) (q \fsl - r \fsl) \right ] + \eta \fsl
\epsilon \fsl (q \fsl - r \fsl) - (\epsilon \eta) (q \fsl - r \fsl) +
2 (\eta q) \epsilon \fsl \right \} L, \label{wo}\\
& &\hspace*{-1cm} \kappa(s) = {1 \over 2} +
{Q_0(s) \over s}, \label{kappa} \\
& &\hspace*{-1cm}s = {2 q r \over m_b^2},  \label{sxi} 
\end{eqnarray}
$Q_u =2/3$ and the function $Q_0(s)$ defined by
\beq \label{q0}
Q_0(s) = \int_0^1 {dx \over x} \ln (1 - s x + s x^2 - i \delta), 
\eeq
with $\delta$ positive infinitesimal. The integration in (\ref{q0}) can
be performed analytically, yielding $Q_0(s)$ as given in eq. (9) of
Ref. \cite{ali2}.  Note that all UV divergences cancel to yield an
UV-finit result. Note also that (\ref{t7}), (\ref{t8}) and (\ref{t2})
are in complete agreement with Ref.\ \cite{ali2} after some
rearrangements in the Dirac structure using the Dirac equation.

\subsection {Penguin operators: $\hat O_3$ -- $\hat O_6$}

Evaluating diagrams containing penguin operators, one must take into
account two types of possible insertions: the somewhat natural one of
Fig.\ 3d, but also the one that leads to diagrams like Fig.\ 3e. The
latter stems from terms in the second current of the operator
containing the $b$ or the $s$ quark. But as it turns out, only one
insertion for every operator survives due to colour structure.

In a calculation analogous to the previous one and without making a
Fierz transformation we obtain:
\begin{eqnarray}
\label{t3bist6}
T_3 &=& Q_d \left [ \kappa_b +
\kappa_s \right ] W_2, \\
T_4 &=& \left \{ {1 \over 6} + Q_d \left [\kappa_b +
\kappa_s  \right ] + Q_u \kappa_c \right \} W_2, \\
T_5 &=&  Q_d \left ( m_b W_5^b R + m_s W_5^s 
 L \right ), \\
T_6 &=& - T_4,
\end{eqnarray}
where we have used the abbreviation $\kappa_q \equiv \kappa \left
({m_b^2 \over m_q^2} s \right )$. $W_2$, $\kappa(s)$, and $s$ are
defined in (\ref{wo}), (\ref{kappa}) and (\ref{sxi}), respectively,
and
\begin{eqnarray}
\nonumber
W_5^q &=&  {1 \over m_q^2} \left [ \eta \fsl r \fsl \epsilon
\fsl q \fsl + (q r) \eta \fsl \epsilon \fsl + (\eta \epsilon) r \fsl q
\fsl - (\epsilon r) \eta \fsl q \fsl - (\eta q) r \fsl \epsilon \fsl
 \right ] (1 - 2 \kappa_q ) \\ & & + \; 4 \left [ (\eta \epsilon) - {
 (\epsilon r) (\eta q) \over q r } \right ] \kappa _q.
\end{eqnarray}

There is one more point to be considered: in the case of the penguin
operators $\hat O_5$ and $\hat O_6$, there are also non-vanishing
diagrams in which at least one photon (or gluon) is radiated from an
external leg. But as in $b \rar s \gamma, sg$ the resulting amplitudes
are proportional to either $T_7$ or $T_8$. Thus the same redefinition
of the Wilson coefficients as in $b \rar s \gamma, sg$ absorbs these
contributions into the tree-level amplitudes $T_7$, $T_8$. This is why
we have introduced the quantities $\tilde C_i$ as defined in
(\ref{effwilson})\footnote{ Strictly speaking, these diagrams
contribute only in the NDR scheme; in the HV scheme they vanish and
therefore in the latter $\tilde C_i=C_i$. This difference is
compensated by a corresponding scheme dependence of the leading order
anomalous dimension matrix.}.

\section{Spectrum and decay rate}

We now proceed to calculate some physical observables from the
amplitude $M^{brems}$ combined with $M_7^{virt}$.

\subsection{Squared and summed amplitude}

Summing the squared amplitude $\vert M^{brems} \vert ^2$ over spins,
polarizations and colours is a straightforward procedure. First of all,
we can write
\beq \label{betragsquadrat}
{\vert {M^{brems}} \vert ^2}_{\Sigma} = {1 \over 6}
\sum\limits_{spin, pol., col.} \vert {M^{brems}} \vert ^2 = {2  \over
3} m_b^4 \vert V \vert ^2
\sum\limits_{i,j=1 \ldots 8 \atop j \ge i} \tilde C_i(\mu)
\tilde C_j(\mu) M_{ij},
\eeq
where the factor 1/6 stems from averaging over spin and colour of the
incoming b quark and V is given in (\ref{V}).  Introducing the
kinematical variables
\beq \label{kinvar}
s = {2 q r \over m_b^2}, \; t = {2 p r \over m_b^2}, \; u = {2 p q
\over m_b^2}
\eeq
and again using the abbreviation $\kappa_q \equiv \kappa \left ({
m_b^2 \over m_q^2} s \right )$, the contributions
from the various operators can be expressed in the limit $m_s=0$ as

\begin{eqnarray}
\nonumber
& & M_{22} = {8 \over 9} \vert \kappa_c \vert^2 (1-s) ,\\
\nonumber
& & M_{23} = - {8 \over 9} {\rm Re} \left [\kappa_c^{\ast}({1 \over 2}
 + \kappa_b) \right ] (1-s), \\
\nonumber
& & M_{24} = - M_{26} = { 8 \over 9} {\rm Re} \left [
\kappa_c^{\ast}(2 \kappa_c -
\kappa_b) \right ] (1-s), \\
\nonumber
& & M_{25} = - {4 \over 9} {\rm Re} \left[ \kappa_c^{\ast}(1-2
\kappa_b) \right ] (1-s) s, \\
\nonumber
& & M_{27} = - 3 M_{28} = - {4 \over 3} {\rm Re} (\kappa_c) s, \\
\nonumber
& & M_{33} = {1 \over 9} \vert {1 \over 2} + \kappa_b \vert ^2 (1-s),
\\
\nonumber
& & M_{34} = - M_{36} = - {4 \over 9} {\rm Re}  \left [({1 \over 2} +
\kappa_b^\ast) (2 \kappa_c
- \kappa_b) \right ] (1-s), \\
\nonumber
& & M_{35} = {2 \over 9} {\rm Re} \left [ ( {1 \over 2} +
\kappa_b)(1-2 \kappa_b^\ast) \right ] (1-s) s, \\ 
\nonumber
& & M_{37} = - 3 M_{38} = {2 \over 3} {\rm Re} ({1 \over 2} +
\kappa_b) s,\\
\nonumber
& & M_{44} = M_{66} = {1 \over 9} \vert 2 \kappa_c - \kappa_b \vert ^2
(1-s),\\
\nonumber
& & M_{45} = - M_{56} = - {2 \over 9} {\rm Re} \left [ (2
\kappa_c-\kappa_b)(1-2 \kappa_b^\ast) \right ] (1-s) s, \\
\nonumber
& & M_{47} = - M_{67} = -3 M_{48} = 3 M_{68} = - {2 \over 3} {\rm Re}
(2
\kappa_c - \kappa_b) s,\\
\nonumber
& & M_{55} = {1 \over 9} \left [ 16 \vert \kappa_b \vert ^2 + \vert (1
 - 2 \kappa_b) s + 4 \kappa_b \vert ^2 \right ] (1-s), \\
\nonumber
& & M_{57} = -3 M_{58} = - {8 \over 3} {\rm Re} (\kappa_b) s, \\
& & M_{78} =  {1 \over 3} \left ( 1+ {2 \over tu} \right ) s.
\label{betragsquadrat1}
\end{eqnarray}
For $m_s \ne 0$ the corresponding expressions become quite lengthy. The
complete formulas for this case can be found in the appendix.

There are still two contributions missing. $M_{77}$ is the one which
will by far dominate the photon spectrum (at least near the endpoint),
so in this case it may be interesting to know the full formula
(i.e. $m_s \ne 0$), which reads
\begin{eqnarray}
\nonumber
& &M_{77} = (1+\rho) \left \{ (1-\rho) M_{77}^{(1)} - 2 M_{77}^{(2)} \right
\},
\\
\nonumber
& &M_{77}^{(1)} = {1 \over \bar t} \left [ 1+ \bar u + {2 \bar u (\bar
u -2)
\over 1- \bar u} \bar t + {2 \bar u - 1 \over 1 - \bar u} \bar t ^2
\right ], \\
& &M_{77}^{(2)} = {1 \over \bar t^2} \left [1 - {1+\rho \over 1- \bar u}
\bar t + { \rho
\over (1- \bar u)^2} \bar t^2 \right ],
\label{m77}
\end{eqnarray}
where we have introduced the rescaled kinematic variables $\bar t =
t/(1-\rho),\; \bar u = u/(1-\rho)$, and $\rho = m_s^2 / m_b^2$ from
(\ref{virt2}). In writing (\ref{m77}), we have split $M_{77}$ into
an infrared safe part $M_{77}^{(1)}$ and an infrared divergent part
$M_{77}^{(2)}$. From the amplitude (\ref{t8}) it should then be
obvious that the quantity $M_{88}/{Q_d^2}$ is given by the above
expression for $M_{77}$ if one interchanges $\bar u$ with $\bar t$.

\subsection{The infrared finite parts}

As stated above, in the region of experimental interest all
contributions $M_{ij}$ to the squared amplitude yield a IR finite
spectrum and decay rate, with exception of $M_{77}$. For those finite
contributions, we therefore can do all phase space integrations in 4
dimensions. Note that from conservation of energy-momentum we have the
relation $u+t-s=1-\rho$ and the kinematic boundaries $u
\in [0,1-\rho], \; t \in [1-u-\rho, 1- {\rho \over 1-u}]$. Furthermore,
in the rest frame of the decaying $b$ quark the variables $u$ and $t$
become the energy of the emitted photon and gluon, respectively,
measured in units of $m_b/2$.  Splitting the spectrum into an infrared
finite and an infrared singular part,
\beq \label{spektrum1}
{d \Gamma^{brems} \over du} = {d \Gamma_F \over du} + {d
\Gamma_7^{brems} \over du},
\eeq
one therefore obtains trivially for the former
\beq \label{spektrumfinit}
{d \Gamma_F \over d u} = \sum\limits_{j \ge i \atop ij \ne 77}
{d \Gamma_{ij} \over d u} = {2 \alpha_s \over 3 \pi} \Gamma_0
\sum\limits_{j \ge i \atop ij \ne 77}
\tilde C_i(\mu) \tilde C_j(\mu)
\int\limits_{t_{min}}^{t_{max}} dt M_{ij} (t,u),
\eeq
with
\beq
\Gamma_0 = {\vert V_{tb} V_{ts}^{ \ast} \vert ^2 G_F^2 \alpha \over 32
\pi^4} m_b^5.
\eeq
Regrettably, due to the presence of the functions $\kappa(s)$ in the
expressions (\ref {betragsquadrat1}) for $M_{ij}$ the integrations in
(\ref{spektrum1}) cannot be done analytically, so we postpone the
numerical analysis of (\ref{spektrum1}) to sections 5.1 (spectrum) and
5.2 (decay width).

\subsection{The infrared divergent part}

Integration of $\vert M_7 \vert ^2$ yields a photon spectrum which
contains a non-integrable singularity as the photon energy reaches the
kinematical endpoint, $E_{\gamma} \rar {m_b \over 2} (1-\rho)$. To
regularize this divergence, we will do the whole phase space
integration in $D=4-2 \epsilon$ dimensions. The singularity will then
manifest itself in a part of the decay width $\Gamma_7^{brems} =
\int d\Gamma_7^{brems}$ proportional to
  $1/ \epsilon$. This singularity will cancel if we add the equivalent
singular contribution $\Gamma_7^{virt}$ stemming from the virtual
gluons, leaving us with an unambiguous finite result $\Gamma_7 =
\Gamma_7^{brems}+\Gamma_7^{virt}$. Let us now calculate $\Gamma_7$.

In $D$ Dimensions, the Lorentz Invariant Phase Space (LIPS) is given by
\beq
d \Phi(p) = {d^{D-1}p \over (2 \pi)^{D-1} 2 p_0},
\eeq
and accordingly the differential decay width for an n-body decay with
transition amplitude M reads
\beq
d \Gamma = {1 \over 2 m} (2 \pi)^D \delta^{(D)} \left ( p -
\sum\limits_{i=1}^n p_i \right ) \vert M
\vert ^2 \prod\limits_{i=1}^n
d \Phi(p_i).
\eeq
Specializing this to the case $b \rar s \gamma g$ and performing the
delta function as well as the angular integrations (note that our
squared and summed amplitude $\vert M^{brems} \vert _{\Sigma}^2$
(\ref{betragsquadrat}) only depends on the energies of the particles
in the final state), one obtains
\beq
d\Gamma_7^{brems} = (1-\rho)^2 \tilde C_7^2(\mu) {2 \alpha_s \over 3 \pi}
\Gamma_0 {(1-\rho)^{-4 \epsilon}
\left ( {8 \pi \mu ^2 \over m_b^2} \right )^{2 \epsilon}  \over \Gamma
(2 - 2 \epsilon) } \left [ \bar t \bar u (1-z(\bar t, \bar u)^2)
\right ]^{-2 \epsilon} M_{77}(\bar t, \bar u) \; d \bar t \; d \bar u,
\eeq
where we have defined the function
\beq
z(\bar t, \bar u) = {2 \over 1-r} \left ( \bar u + \bar t - 1 \over
\bar u \bar t \right ) -1,
\eeq
which is actually the cosine of the angle between $\vec q$ and $\vec
r$ in the rest frame of the decaying $b$ quark.  The crucial point to
be observed now is that even in the first integration over $\bar t $
(which yields no divergences) one has to hold terms of order
$\epsilon$, because they may multiply later on with $1/\epsilon$ poles
of the $\bar u$ - integration and so give finite contributions to the
total decay width.

After carrying out the integration over $\bar t$, we obtain the
following expression for the photon spectrum to $O(\epsilon)$:

\beq \label{spektrum7}
{d \Gamma_7^{brems} \over d \bar u} = (1+\rho)(1-\rho)^3 \tilde C_7^2(\mu)
{2 \alpha_s \over 3 \pi}
\Gamma_0  C_{\epsilon} \left [ S^{(1)} - {2 S^{(2)}
\over (1- \bar u)^{1+2 \epsilon } }\right ],
\eeq
with
\begin{eqnarray}
\nonumber
C_{\epsilon} &=& {(1-\rho)^{-4 \epsilon} \left ( {4 \pi \mu^2 \over
m_b^2}
\right ) ^{2 \epsilon} \over \Gamma(2-2 \epsilon) } \bar u^
{-2 \epsilon}, \\
\nonumber
S^{(1)} &=& {1 \over 2} (1-\rho) {\bar u (1- \bar u) (2 \bar u -1)
\over (1-u)^2 } + {1 \over 2} (1-\rho) {\bar u (2 \bar u ^2 - 5 \bar u -
1) \over 1-u } - (1 + \bar u) \ln (1-u), \\
\nonumber
S^{(2)} &=& S^{(2)}_a + \epsilon S^{(2)}_b, \\
\nonumber
S^{(2)}_a &=& \left ( 1 + {\rho \over 1-u} \right ) \bar u + {1+\rho
\over 1-\rho} \ln (1-u), \\
S^{(2)}_b &=& {\rho \over 1-u} \left [ 2 + \ln (1-u) \right ] \bar u -
2 {1-u \over 1-\rho} \ln (1-u) + {1+\rho \over 1-\rho} \left [ {1 \over 2} \ln
^2(1-u) - 2 \; {\rm Li_2}(u) \right ],
\end{eqnarray}
$\rho = m_s^2/m_b^2$ and ${\rm Li_2}(x)= - \int_0^1 {dt \over t} \ln
(1-x t)$ is the Spence function (or dilogarithm).  From this, we can now
accomplish our final goal, the integration over $\bar u$. Since our
results agree entirely with (29) and (30) of Ref. \cite{ali2}, we will
not give them here explicitly.

Then one has to go through the same steps with $M_7^{virt}$: square
it, sum over spins and polarizations, do the phase space integration
in $D$ dimensions. The result will contain a singular term that cancels
the one of $\Gamma_7^{brems}$. In fact these calculations are much
simpler than the one sketched above, because we only have to struggle
with a two-body decay. After some work, we arrive at the following
final result\footnote{The attentive reader will perhaps feel uneasy
about the fact that in order to obtain the following result, we
apparently have calculated $\Gamma_7^{virt}$ with the effective Wilson
coefficient $\tilde C_7$ instead of $C_7$. Of course the full two-loop
matrix elements of $\hat O_5$ and $\hat O_6$ are not proportional to
$M_7^{virt}$.  Nevertheless, due to cancellation of infrared
divergences they must contain at least a part proportional to
$M_7^{virt}$. The other parts have to be considered in the full
calculation of two-loop $b \rar s
\gamma$ matrix elements, which is not the aim of this paper.}
for the total decay width via $\hat O_7$:
\beq \label{spektrum7tot}
\Gamma_7 = \Gamma_7^{(0)} + \delta \Gamma_7,
\eeq
with
\beq \label{spektrum7tot1}
\Gamma_7^{(0)} = (1+\rho) (1-\rho)^3 \tilde C_7(\mu)^2 \left
( {\overline m_b(m_b)
\over m_b} \right ) ^2 \Gamma_0
\eeq
and
\begin{eqnarray}
\nonumber
\delta \Gamma_7 &=& {2 \alpha_s \over 3 \pi}
\left \{ {4 \over 3} {2 - 7 \rho + 2 \rho^2 \over (1-\rho)^2} - {\rho
 (7 - 8 \rho + 5 \rho^2) \over (1-\rho)^3} \ln \rho -
 4 \ln (1-\rho) \right . \\ & &
\left . + {1+\rho
\over 1-\rho} \left [ 4 \; {\rm Li_2} (\rho) - {2 \over 3} \pi^2 + 2 \ln \rho
\ln (1-\rho) \right ] + 4 \ln {m_b^2
\over \mu ^2 } \right \} \Gamma_7^{(0)}.  \label{spektrum7tot2}
\end{eqnarray}
In deriving (\ref{spektrum7tot}) -- (\ref{spektrum7tot2}) we have expanded the running quark
mass $\overline m_b(\mu)$, which occurs from (\ref{t7virtR}) in the
matrix element $M_7^{virt}$, around $\mu = m_b$,
\beq
\overline m_b(\mu) = \overline m_b(m_b) \left (1 + 
{2 \alpha_s(m_b)
\over  \pi} \ln {m_b \over \mu} \right ),
\eeq
and we have also made use of the equivalence of $\overline m_b(m_b)^2$
and the squared pole mass $m_b^2$ up to corrections of order
$\alpha_s(m_b)$.  In the limit $m_s=0$ (\ref{spektrum7tot}) looks much
nicer:
\beq \label{gamma7withoutmass}
\Gamma_7 = \tilde C_7(\mu)^2 \left [ 1 + {2 \alpha_s \over 3
\pi}
 \left ( {8 - 2 \pi^2 \over 3} + 4 \ln {m_b^2 \over \mu^2} \right )
\right ] \left ( {\overline m_b(m_b) \over m_b} \right )^2 \Gamma_0.
\eeq
We now hasten to undertake a critical discussion of all these results.

\section{Numerical analysis}

\subsection{Photon spectrum}

If we neglect all non-perturbative effects, what does the photon
spectrum due to bremsstrahlung actually look like? The answer depends
on the experimental setup: Let us assume the resolution of the photon
detector is ${m_b \over 2} \Delta$ so that in the endpoint region we
cannot discriminate photons within the energy interval $E_\gamma^{max}
- {m_b
\over 2} \Delta < E_\gamma < E_\gamma^{max}$, with $E_\gamma^{max} =
{m_b^2-m_s^2 \over 2 m_b}$ the maximum energy of the emitted
photon. If we now define the function
\beq
\Gamma_{tot}(u_0) = \Gamma_7 - \int_0^{u_0} {d \Gamma_7^{brems} \over
du} du + \int_{u_0}^{1-\rho} {d \Gamma_F \over du} du,
\eeq
which is the total decay width including all photons with an energy
$E_\gamma \ge {m_b \over 2} u_0$, then the photon energy spectrum is
given by
\beq \label{photonspektrum}
{d \Gamma_\Delta \over d u} = \left \{ 
\begin{array}{cl} {d \Gamma^{brems} \over  du}, & u < 1 - \rho - \Delta \\
{1 \over \Delta} \Gamma_{tot} \left ( 1-\rho-\Delta \right ), & 1 - \rho -
\Delta < u < 1 - \rho \\ 0, & u > 1 - \rho
\end{array} \right . 
\eeq

Here, $d\Gamma^{brems}/du$ is the perturbative bremsstrahlung
spectrum which we calculated in Sec.\ 4, cf.\ eqs.\ (\ref{spektrum1}),
(\ref{spektrumfinit}) and (\ref{spektrum7}). We have plotted the
resulting photon spectrum in Fig.\ 4,
\begin{figure}[t]
\centerline{
\rotate[r]{
\epsfysize=15cm
\epsffile{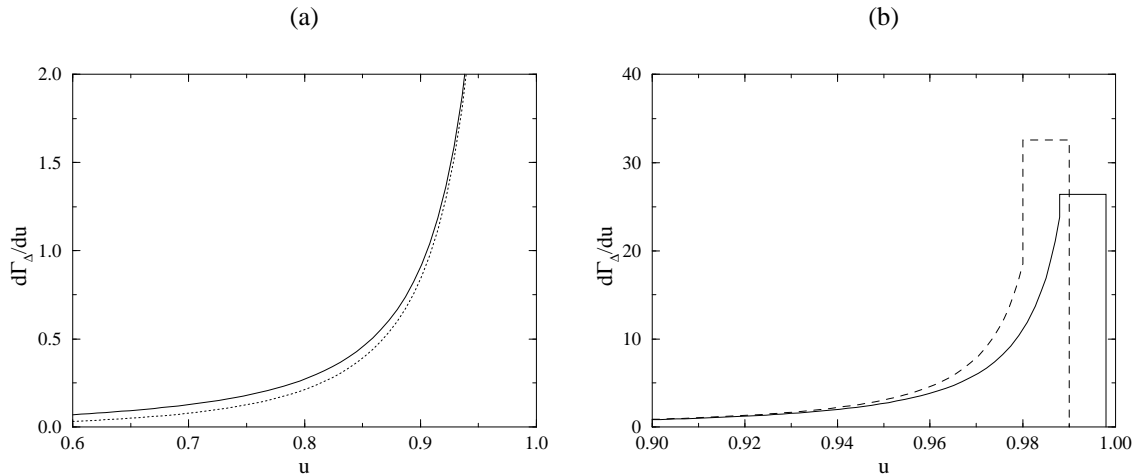}}
}
\caption{\small Two different views on the $B \rar X_s \gamma$ photon
energy spectrum due to gluon bremsstrahlung, eq.\ (\ref
{photonspektrum}). The spectrum is given in units of $\vert \tilde
C_7(m_b) \vert ^2 \Gamma_0$, the leading order decay width with
$m_s=0$. In both plots, the solid curve corresponds to the full
operator basis and $m_s = 200\;{\rm MeV}$. (a) shows the region
$0.6<u<0.95$, where $u$ is the energy of the emitted photon in units
of $m_b/2$. The dotted curve represents the contribution from the
operator $\hat O_7$ alone, which is obviously the dominant one. (b)
takes a closer look at the endpoint region, where the spectrum is
sensitive to the mass of the strange quark. Here, the dashed curve
depicts the case $m_s=500\;{\rm MeV}$. }
\end{figure}
where we have used $m_s = 200 \; {\rm MeV}$, $\Delta=25\;{\rm MeV}$
and the Wilson coefficients $\tilde C_i(m_b)$ as given and discussed
in the following section. Though we will not dwell upon any further
details, a few comments may be useful:

First of all, if we consider the limit $\Delta \rar 0$, up to what
minimal resolution $\Delta_{crit}$ the spectrum (\ref{photonspektrum})
can be trusted? Clearly, this is a question of the reliability of
perturbation theory. We can estimate $\Delta_{crit}$ if we consider
the case where the pure O($\alpha_s$) bremsstrahlung contributions
from the region $u < 1 - \rho -
\Delta$ become of the same order of magnitude as the
contributions from the peak in the region $1-\rho-\Delta < u < 1 - \rho$:
\beq \label{deltacrit}
\int\limits_0^{1-\rho-\Delta_{crit}} {d \Gamma^{brems} \over
du} du \; \simeq \; \Gamma_{tot} \left (1-\rho-\Delta_{crit} \right ) .
\eeq

This is, because if the l.h.s.\ of (\ref{deltacrit}) becomes greater
than the r.h.s., then the O($\alpha_s$) brems-strahlung corrections
exceed the two-body decay peak so that perturbation theory is no
longer valid. If one wishes to explore the case $\Delta <
\Delta_{crit}$, one will have to resort to some Sudakov-type partial
resummation of the most singular terms (``exponentiation'') in order
to obtain a reliable spectrum. However, numerically we get from
(\ref{deltacrit}), with $m_s = 200 \; {\rm MeV}$, $\Delta_{crit}
\simeq 0.015$, so that an infrared improvement remains
unquestionably necessary only in the last, say, 30--40 MeV near the
kinematical endpoint. This is well below any experimental resolution,
and moreover it will presumably be of no importance for the lower
moments of the spectrum \cite{kapustin}. Therefore, we suggest to take
(\ref{photonspektrum}) with $\Delta \simeq 0.015$ as the perturbative
photon spectrum and as the input for the inclusion of non-perturbative
effects.

From Fig.\ 4a one sees that the spectrum is completely dominated by
the contribution stemming from the Operator $\hat O_7$ (solid vs.\
dotted line). We can express this observation in a more quantitative
way, if we define the (normalized) moments of (\ref{photonspektrum})
as
\beq \label{moments}
M_n(u_0,\Delta) = {1 \over \Gamma_{tot}(u_0)} \int_{u_0}^{\infty} u^n 
{d \Gamma_\Delta \over du} \, du, \; \;  n \in \Nfat.
\eeq
For instance, we obtain thereby for the first moment $M_1(0.6,0.015) =
0.961$ if we include contributions from all operators, and
$M_1(0.6,0.015) = 0.958$ with $C_i(\mu) \equiv 0$ for $i \ne 7$. These
dominance of $\hat O_7$ is a quite welcome feature, because
$d\Gamma^{brems}_7/du$ is the only contribution which we can calculate
analytically (without any numerical integration).

The moments (\ref{moments}) are also useful to illustrate the
dependence of the spectrum on the mass of the strange quark. Varying
$m_s$ between 200 and 500 MeV (solid vs.\ dashed line in Fig.\ 4b),
$M_1(0.6,0.015)$ changes from 0.961 to 0.952. So the moments are {\em
decreased} by increasing values of $m_s$, in contrary to the result
found in Ref.\ \cite{kapustin}. This is primarily due to our slightly
different normalization of the moments, eq. (\ref{moments}), which we
have chosen because the so-defined moments can in principle be
extracted from experiment without knowledge of $m_s$. On the other
hand, it is also due to our different treatment of the infrared
divergent endpoint region: whereas Kapustin and Ligeti trusted the
perturbative bremsstrahlung spectrum up to arbitrary small $\Delta$,
with (\ref{photonspektrum}) we rely on first order perturbation theory
only where it is clearly valid. Since the $m_s$-dependence of the
spectrum manifests itself mainly in the endpoint region, we regard our
method as the more appropriate one.

\subsection{Decay width}

Taking into account the discussion of Sec.\ 2 it should be perfectly
clear that from our calculations we cannot present any definite
numbers for the total decay width (or, equivalently, the branching
ratio). What we can do is, however, to list the complete
bremsstrahlung contributions to the decay rate. When we will know all
the remaining parts of the NLO analysis, especially the two-loop $b
\rar s \gamma$ matrix elements of the four-quark operators, we may
then simply put together all these numbers to get some final answer.

We will give our numerical results in units of $\vert \tilde C_7(m_b)
\vert ^2 \Gamma_0$, which is the leading order decay width if
 one neglects the mass of the strange quark. Therefore, our only free
parameter is $\rho=m_s^2/m_b^2$. Additionally, we have to assume some
value for $E_{\gamma}^{min}$, the minimum photon energy we can see with
the detector. For some choices of $\rho$ and $E_{\gamma}^{min}$, the
resulting bremsstrahlung contributions to the total decay width
calculated from (\ref{spektrumfinit}) and (\ref{spektrum7tot}) are shown
in Table 1.
\begin{table}[t]
\begin{center}
\begin{tabular}{|c|r||ccccccc|c|}
\hline
$\rho$ & $E^{min}_\gamma$ & $\Gamma_{7}$ & $\Gamma_{22}$ & $\Gamma_{27}$
& $\Gamma_{28}$ & $\Gamma_{78}$ & $\Gamma_{88}$ &
$\sum\limits_{\rm{all \; others}}
\Gamma_{ij}$ &
$\Gamma_{tot}$
\\ 
\hline \hline
& 1.92 GeV & $66.92$ & $0.94$ & $-0.09$ & $0.03$ & $0.16$ & $0.02$ &
 0.00 & $67.98$ \\
\cline{2-10}
0.01 & 1.44 GeV & $68.83$ & $1.85$ & $-0.27$ & $0.08$ & $0.32$ &
$0.09$ & 0.01  & $70.91$\\
\cline{2-10}
& 0 GeV & $69.16$ & $2.48$ & $-0.48$ & $0.15$ & $0.48$ & - & 0.02 & -
\\
\hline
& 1.92 GeV & $67.54$ & 1.14 & $-0.08$ & $0.02$ & 0.20 & 0.06 & 0.01 &
 $68.89$ \\
\cline{2-10}
0.002 & 1.44 GeV & $69.09$ & $2.12$ & $-0.27$ & $0.06$ & $0.37$ & 0.18
& 0.01 & $71.52$ \\
\cline{2-10}
& 0 GeV & $69.40$ & 2.80 & $-0.48$ & $0.10$ & 0.54 & - & 0.02 & - \\
\hline
\end{tabular}
\end{center}
\caption{\small Different contributions to the total decay width $B
\rar X_s \gamma$ due to gluon bremsstrahlung, in percent of
$\vert \tilde C_7(m_b) \vert ^2\,\Gamma_0$, the leading order decay
width with $m_s=0$, as a function of the parameters $\rho=m_s^2 / m_b^2$
and $E_\gamma^{min}$. We have used the leading order effective Wilson
coefficients from Table 2, $m_c = 1.4 \, {\rm GeV}$, $m_b = 4.8 \,
{\rm GeV}$ and $\overline m_b(m_b) = 4.4 \, {\rm GeV}$.  The last
column shows the sum of all the different contributions.}
\end{table}

In our analysis we have used the leading order effective Wilson
 coefficients $\tilde C_i(\mu)$, see e.g.\ \cite{LO} for explicit
 formulas, as well as the two-loop running coupling $\alpha_s(\mu)$,
 and we set $\mu = m_b$ with $m_b = 4.8 \, {\rm GeV}$ being the pole mass of
 the $b$ quark. The resulting numerical values for the Wilson
 coefficients $\tilde C_i(m_b)$ are displayed in Table 2, where they
 have been computed with the parameters $M_W = 80.2\, {\rm GeV}$,
 $\overline m_t(m_t) = 170 \, {\rm GeV}$ and $\alpha_s(M_Z) = 0.117$.
\begin{table}[t]
\begin{center}
\begin{tabular}{|c|c|c|c|c|c|c|c|}
\hline
$\tilde C_1$ & $\tilde C_2$ & $\tilde C_3$ & $\tilde C_4$ & $\tilde
C_5$ &
 $\tilde C_6$ & $\tilde C_7$ & $\tilde C_8$ \\
\hline
-0.242 & 1.104 & 0.011 & -0.025 & 0.007 & -0.030 & -0.309 & -0.147 \\
\hline
\end{tabular}
\caption{\small Effective Wilson 
coefficients $\tilde C_i(\mu)$ at the scale $\mu = m_b = 4.8
\, {\rm GeV}$, in the approximation of leading logarithms. These
coefficients were evaluated with the two-loop $\alpha_s(\mu)$ and the
parameters $M_W = 80.2\,{\rm GeV}$, $\overline m_t(m_t)=170\,{\rm
GeV}$, $\alpha_s(M_Z) = 0.117$.}
\end{center}
\end{table}

Let us now discuss some interesting features of Table 1:

\bigskip
\noindent
{\em (i) Contributions from different operators}

\noindent
The dominant contribution to $\delta\Gamma_{tot}=
\Gamma_{tot}-\Gamma_7^{(0)}
$comes from the operator $\hat O_7$ alone, all other contributions
amount to only 15\% of this one. However, this statement has to be
taken with a grain of salt, because $\hat O_7$ is the only operator
where we have included the virtual corrections. Contributions from the
penguin operators $\hat O_3$ -- $\hat O_6$ (last but one column) are
completely negligible as far as they are not included in the
definitions of the effective Wilson coefficients. Whereas the virtual
corrections lower the decay rate significantly, bremsstrahlung tends
to enhance it slightly.

\bigskip
\noindent
{\em (ii) Dependence on $m_s$ and $E_{\gamma}^{min}$}

\noindent
Since $m_s^2 \ll m_b^2$, from the outset we did not expect the mass of
the strange quark to make any sizeable effect in $\Gamma_{tot}$, and
we used $m_s$ therefore mainly as regulator of collinear
divergences. Indeed it turns out that varying $m_s$ between 0 and 500
MeV leads to a change of only 0.9\% in $\Gamma_{tot}$. Here, we should
note in particular that $\Gamma_7=\Gamma_7^{brems}+\Gamma_7^{virt}$
and all bremsstrahlung contributions $\Gamma_{ij}$ with $ij \ne 77$
are well defined in the limit $m_s \rar 0$, except for
$\Gamma_{88}$. This divergent behaviour of $\Gamma_{88}$ has recently
been discussed in Ref. \cite{kapustin2}, but since, from Table 1,
$\Gamma_{88}$ is in the relevant part of the spectrum and for sensible
values of $m_s$ of no numerical importance, this observation is of no
great relevance here.

The dependence on $E_\gamma^{min}$ can be interpolated from Table 1.
Note that values $E_\gamma^{min} < 1.4 \; {\rm GeV}$ are not
realistic. Above that, if one is interested in the limit
$E_\gamma^{min} \rar 0$, one has to deal with the infrared (soft
photon) divergence due to $\hat O_8$, cf. Ref. \cite{ali1, kapustin2}.

\bigskip
\noindent
{\em (iii) Dependence on the renormalization scale $\mu$}
 
\noindent
As mentioned in Sec.\ 2, an important offspring of the complete NLO
calculation should be a considerably reduced renormalization scale
dependence and thus a considerably improved theoretical uncertainty.
From theory we know that in order to perform the correct
resummation of large logarithms the renormalization scale $\mu$ should
be fixed to a value of order $m_b$, but that on the other hand it is
arbitrary provided that $\ln {m_b \over \mu}$ remains a small
parameter. Thus this uncertainty can be estimated if one varies $\mu$
between, say, $m_b/2$ and $2 m_b$. Performing this exercise with the
expression (\ref{gamma7withoutmass}) for $\Gamma_7$, one obtains a
startling large relative change of $+69\%
\atop -38\%$. This is an even larger $\mu$-dependence than the $+28\%
\atop -20\%$ of the pure leading order result. What lesson can
 we learn from this peculiar effect? Of course a very elementary one:
namely, that it is a mistake to expect any improved predictive power
from {\em partial} NLO calculations. Let us explain this in detail.

The $\mu$-dependence of the Wilson coefficients $\tilde C_i(\mu)$ is
governed by the renormalization group equation \cite{burasreview}
\beq \label{RGE}
\mu {d \over d \mu} \tilde C_i(\mu) = \hat \gamma_{ji}(\alpha_s)
\tilde C_j(\mu),
\eeq
where the NLO anomalous dimension matrix $\hat \gamma_{ij}$ is given
by
\beq \label{anomalousdimension}
\hat \gamma (\alpha_s) = \hat \gamma^{(0)} {\alpha_s(\mu)
\over 4 \pi} + 
\hat \gamma^{(1)} \left ( {\alpha_s(\mu) \over 4 \pi} \right ) ^2.
\eeq
Here, $\hat \gamma^{(0)}$ represents the leading order (LO)
contribution, and $\hat \gamma^{(1)}$ the NLO correction to it. $\hat
\gamma^{(0)}$ has been calculated by various authors and is given
e.g.\ in Ref.\ \cite{LO}, and the calculation of $\gamma^{(1)}$ is not
yet completed. From (\ref{RGE}), (\ref{anomalousdimension}) we can
infer that the dominant $\mu$-dependence of $\tilde C_7(\mu)$ near
$\mu = m_b$ is given by
\beq \label{mudependence}
\tilde C_7(\mu) = \tilde C_7(m_b) +  {\alpha_s(m_b) \over 4 \pi}
\gamma_{i7}^{(0)}
\tilde C_i(m_b) \ln{\mu \over m_b} + O \left( \alpha_s^2 \ln{\mu \over
m_b}, \alpha_s^2 \ln^2{\mu \over m_b} \right ).
\eeq
The second term in (\ref{mudependence}) is exactly the one that after
a complete NLO calculation should be cancelled by the
$\mu$-dependence of the $b \rar s \gamma$ matrix elements $\langle
\hat O_i(\mu) \rangle$, since such a calculation should be free of
renormalization scale uncertainties at O($\alpha_s$). From these general
grounds we therefore conclude that the transition amplitude $b \rar s
\gamma$ at the NLO-level,
\beq
M(b \rar s \gamma) = - {4 G_F \over \sqrt 2} V_{tb} V_{ts}^\ast 
\sum\limits_{i=1}^8 C_i(\mu)^{NLO} \langle \hat O_i(\mu)
\rangle,
\eeq
can be written as 
\beq \label{amplitude}
M(b \rar s \gamma) \sim \left [ \tilde C_7(\mu) + {\alpha_s(m_b) \over
4 \pi} \left ( \hat \gamma_{i7}^{(0)} \tilde C_i(m_b) \ln {m_b \over
\mu} + r_i C_i(m_b)
\right ) \right ] \langle O_7 \rangle^{tree},
\eeq
where $r_i$ are functions of the external momenta, spins and
polarizations that contain no $\mu$-dependence, and the last factor
denotes the tree-level matrix element of $\hat O_7$. In
(\ref{amplitude}), the first summand $\tilde C_7(\mu)$ has to be taken
at the NLO level, whereas for the other Wilson coefficients LO is
sufficient, since they are multiplied already with $\alpha_s$. In
order to obtain a prediction for the $\mu$-dependence of the quantity
$\Gamma_7$ which we calculated explicitly in this paper,
cf. eq. (\ref{gamma7withoutmass}), we square (\ref{amplitude}), set
$\tilde C_i(\mu) \equiv 0$ for $i \ne 7$ and get
\beq
\Gamma_7 \sim \tilde C_7(\mu)^2 \left [ 1 + {\alpha_s(m_b) \over 4 \pi}
\left( \gamma_{77}^{(0)}
\ln {m_b^2 \over \mu^2} + \tilde r_7 \right ) \right ],
\eeq
where $\tilde r_7$ is an unpredicted number. With $\gamma_{77}^{(0)} =
{32
\over 3}$ from \cite{LO}, we indeed arrive at the
same coefficient of $\ln {m_b^2 \over \mu^2}$ as found in eq.\
(\ref{gamma7withoutmass}), and cancellation with the $\mu$-dependence
of $\tilde C_7(\mu)$ should work. Obviously, it didn't. This is for
the following reason, which illustrates once more the importance of
operator mixing under renormalization: since we neglected all the
$b \rar s \gamma$ matrix elements $\langle
\hat O_i(\mu) \rangle$ with $i \ne 7$, only that $\mu$-dependence of
$C_7(\mu)$ which is proportional to $\gamma_{77}^{(0)}$, cf.\ eq.\
(\ref{mudependence}), could be cancelled within our calculation.  But
$\gamma_{77}^{(0)} \, \tilde C_7(m_b) <0$ and $\sum\limits_{i \ne 7}
\, \gamma_{i7}^{(0)} \tilde C_i(m_b) > \vert
\gamma_{77}^{(0)} \,  \tilde C_7(m_b) \vert$, so actually
$\tilde C_7(\mu)<0$ is a monotonous growing function of
$\mu$. Consequently, by cancellation of the $\mu$-dependence due to
the anomalous dimension $\gamma_{77}^{(0)}$, this monotonous growth is even
more enlarged instead of being reduced. Such a reduction will only
take place after including the O($\alpha_s$) corrections to all $b
\rar s \gamma$ matrix
elements $\langle O_i(\mu) \rangle$.

\section{Summary}

In summary, we have calculated in this paper the complete
O($\alpha_s$) gluon bremsstrahlung corrections to the decay $B \rar
X_s \gamma$. We confirmed all the results of
Ali and Greub, and additionally we gave in Table 1 a compilation of
the {\em complete} bremsstrahlung contributions to the total decay
width.  We also presented formulas for the photon energy spectrum and
made a simple suggestion how to use this bremsstrahlung spectrum as an
input for the inclusion of non-perturbative effects. We found that in
both cases contributions from the QCD penguin operators $\hat O_3$ --
$\hat O_6$ are entirely negligible, except for those parts of them
which can be taken into account by the use of the effective Wilson
coefficients (\ref{effwilson}) instead of the usual ones. We
investigated the considerably increased renormalization scale
dependence of our results and showed that a reduction of this large
theoretical error will not be possible unless {\em all} $b \rar s
\gamma$ matrix
elements $\langle \hat O_i(\mu) \rangle$ will have been calculated up
to $O(\alpha_s)$. Very recently, the first calculation of such
two-loop matrix elements was indeed performed ß\cite{greub}; as it
should be,
 the results allow for a drastically
reduced renormalization scale error.

Finally, we would like to emphasize the urgent demand for
the calculation of the still outstanding
parts of the complete NLO analysis. Without these calculations, a {\em
consistently} improved prediction for the branching ratio will not be
possible, so we do not include such a prediction in our paper. We hope
that these improvements in theory will be forthcoming soon, and that
it will be possible to compare them with even more improved inclusive
measurements, too.

\section*{Acknowledgements}
I would like to thank Ulrich Nierste and Manfred M\"unz for numerous
helpful discussions, comments and suggestions. I am also very grateful
to Andrzej J. Buras and Mikolaj Misiak for critically reading the manuscript.

\begin{appendix}
\section*{Appendix}
Below we explain how to obtain from eqs.\ (\ref{betragsquadrat1}) the
complete expressions (i.e.\ the case $m_s \ne 0$) for the various
contributions $M_{ij}$ to the squared and summed bremsstrahlung
amplitude (\ref{betragsquadrat}).

\noindent
First of all, in every factor $1/2 + \kappa_b$ and $2 \kappa_c -
\kappa_b$ in (\ref{betragsquadrat1}) replace $\kappa_b$ by $\kappa_b
+ \kappa_s -1/2$.

\noindent
Additionally, if both $i$ and $j$ $\in \{2,3,4,6 \}$: make in the
formulas of (\ref{betragsquadrat1}) the replacement
\beq
(1 -s) \rar (1-\rho)^2 - (1+\rho) s.
\eeq
If $i \ne 5$ and $j=7$: make in the formulas of
(\ref{betragsquadrat1}) the replacement
\beq
s \rar (1+\rho)s +  {2 \rho s^2 \over (s-t) t}.
\eeq
If $i \ne 5,7$ and $j=8$: make in the formulas of
(\ref{betragsquadrat1}) the replacement
\beq
s \rar (1+\rho)s + {2 \rho s^2 \over (s-u) u}.
\eeq
Finally, in the expressions for $M_{25}$, $M_{35}$, $M_{45}$ and
$M_{56}$ make the replacement
\beq
(1-2\kappa_b)(1-s)s \rar \left [ (1-2 \kappa_b) (1-\rho -s) - (1-2
\kappa_s) (1-\rho + s) \right ] s
\eeq
The remaining four expressions read explicitly
\begin{eqnarray}
M_{78} &=&  {s \over 3 \bar t \bar u} \left \{ (1+\rho)(2+ \bar t
\bar u) + {\rho \over 1-\rho } \left [ 1 - {2 (1-\rho) - \bar t \bar u \over
(1- \bar t) (1 - \bar u)} \right ] \bar s \right \}, \\
\nonumber
M_{55} &=& \left \{ 16 \vert \kappa_b \vert ^2 + \vert (1 - 2 \kappa_b
) s + 4
\kappa_b \vert ^2 + \rho \left [ 16 \vert \kappa_s \vert ^2 + \vert {1
\over \rho} (1 -
2 \kappa_s) s + 4 \kappa_s \vert ^2 \right ] \right \} (1-s) \\
& & {}+ 16 \, {\rm Re} \left \{ 8 \rho \kappa_b \kappa_s^\ast + s \left [ 
\kappa_b - 2 (1+\rho) \kappa_b \kappa_s^\ast + \rho \kappa_s^\ast \right ]
\right \}, \\
\label{m57}
M_{57} &=& - { 2 \over 3} {\rm Re} \left \{ 4 (\kappa_b +
\rho \kappa_s) s - [ 4 \rho (\kappa_b +  \kappa_s) + s ( (1-2
\kappa_s) + \rho (1-2 \kappa_b)) ] 
{s^2 \over t (s-t)} \right \}, \\
\label{m58}
M_{58} &=&  {2 \over 9} {\rm Re} \left \{ 4 (\kappa_b +
\rho \kappa_s) s - [ 4 \rho (\kappa_b +  \kappa_s) + s ( (1-2
\kappa_s) + \rho (1-2 \kappa_b))  ] 
{s^2 \over u (s-u)} \right \},
\end{eqnarray}
Here, $s$, $t$ and $u$ are the kinematical variables defined in
(\ref{kinvar}), $\bar x = x/(1-\rho)$ with $x=s,t,u$, $\rho=m_s^2 / m_b^2$
and $\kappa_q \equiv
\kappa \left ( {m_b^2 \over m_q^2} s \right )$ with $\kappa(s)$ from
(\ref{kappa}). $M_{77}$ and $M_{88}$ are already given in (\ref{m77}),
and all other contributions vanish.
\end{appendix}

\end{document}